\documentclass{ifacconf}

\usepackage{graphicx}      % include this line if your document contains figures
\usepackage{natbib}        % required for bibliography

\usepackage[utf8]{inputenc}
\DeclareUnicodeCharacter{2212}{−}
\usepackage{amsmath}
\usepackage{amssymb}
\usepackage{siunitx}
\usepackage{url}

\graphicspath{{figures/}}

\newcommand{\meas}[0]{\mathbf{y}}
\newcommand{\state}[0]{\mathbf{x}}

\newcommand{\prcn}[0]{\mathbf{w}}

\newcommand{\dynA}[0]{\mathbf{A}}

\newcommand{\dynC}[0]{\mathbf{C}}

\newcommand{\dynQ}[0]{\mathbf{Q}}

\newcommand{\dynN}[0]{\mathbf{N}}
\newcommand{\dynEta}[0]{\boldsymbol{\eta}}
\newcommand{\ddt}[0]{\frac{\mathrm{d}}{\mathrm{d}t}}
\newcommand{\ddtau}[0]{\frac{\mathrm{d}}{\mathrm{d}\tau}}
\newcommand{\cov}[0]{\boldsymbol{\Sigma}}

\newcommand{\dynParams}[0]{\mathbf{p}}

\newcommand{\Reals}[0]{\mathbb{R}}
\newcommand{\Imaginaries}[0]{\mathbb{C}}

\newcommand{\fwrdstate}[0]{\overrightarrow{\state}}
\newcommand{\backstate}[0]{\overleftarrow{\state}}

\newcommand{\fwrdcov}[0]{\overrightarrow{\cov}}

\newcommand{\backcov}[0]{\overleftarrow{\cov}}

\newcommand{\modeltimetrace}[0]{\mathcal{M}}
\newcommand{\meastimetrace}[0]{\mathcal{Y}}
\newcommand{\paramsSet}[0]{\mathcal{P}}
\newcommand{\impulseDirection}[0]{\mathbf{l}}

\newcommand{\denmatr}[0]{\boldsymbol{\hat{\rho}}}
\newcommand{\hamil}[0]{\hat{H}}
\newcommand{\sopdet}[0]{\mathcal{D}}
\newcommand{\sopsto}[0]{\mathcal{H}}
\newcommand{\meffi}[0]{\eta}
\newcommand{\mrate}[0]{\kappa}
\newcommand{\measop}[0]{\hat{L}}
\newcommand{\sympmatr}[0]{\boldsymbol{\Omega}}
\newcommand{\quadvec}[0]{\mathbf{x}}
\newcommand{\quadop}[0]{\hat{\mathbf{x}}}
\newcommand{\quadcov}[0]{\boldsymbol{\Sigma}}

\newcommand{\hamilMatr}[0]{\mathbf{H}}
\newcommand{\hamilVec}[0]{\mathbf{d}}
\newcommand{\collapseVec}[0]{\mathbf{b}}

\newcommand{\fnoise}[0]{S_{\text{FF}}}
\newcommand{\mnoise}[0]{S_{\text{nn}}}

\newcommand{\dif}[0]{\mathrm{d}}

\newcommand{\matrsqr}[1]{#1^{\frac{1}{2}}}
\newcommand{\trans}[1]{{#1}^{\mathrm{T}}}
\newcommand{\trace}[1]{\text{Tr}\left[#1\right]}

\newcommand{\adj}[1]{#1^{\dagger}}
\newcommand{\Exp}[1]{\mathbb{E}\left[#1\right]}
\newcommand{\realPart}[1]{\text{Re}\left(#1\right)}
\newcommand{\imagPart}[1]{\text{Im}\left(#1\right)}
\newcommand{\diag}[1]{\text{diag}\left(#1\right)}

\newcommand{\Dataset}[0]{\mathcal{Z}}
\begin{document}
\begin{frontmatter}

\title{Optimal State Preparation for Impulse Estimation in Gaussian Quantum Systems} 

\thanks[footnoteinfo]{This research was funded in part by the Austrian Science Fund (FWF) [10.55776/COE1; 10.55776/PAT9140723; 10.55776/P36236].}

\author[ACIN]{K. Schmerling} 
\author[ACIN,AIT] {A. Kugi}
\author[ACIN]{A. Deutschmann-Olek} 
%\author[Third]{Third C. Author}

\address[ACIN]{Automation and Control Institute, TU Wien, Vienna, Austria}
   %Vienna, Austria (e-mail: schmerling@acin.tuwien.ac.at).}
%\address[Second]{Automation and Control Institute Technical University of Vienna, 
%   Vienna, 1040 Austria (e-mail: author@lamar. colostate.edu)}
\address[AIT]{AIT Austrian Institute of Technology, Vienna, Austria}

\begin{abstract}                % Abstract of 50--100 words
%We develop a control-based framework to enhance impulse detection sensitivity in classical and quantum harmonic oscillators by exploiting non-equilibrium squeezing dynamics. Using the Kalman–Bucy formalism for linear Gaussian systems, we cast the minimization of impulse estimation uncertainty as a nonlinear optimal control problem over time-dependent system parameters. The method dynamically shapes covariance evolution to maximize information gain at a known impulse time. Applied to nanomechanical resonators and optically levitated nanoparticles, optimal parametric modulation of stiffness or optical power reduces estimation variance by up to a factor of two compared with conventional rectangular driving. The framework provides a general approach for dynamic noise squeezing and precision enhancement in continuously monitored sensing platforms.
We present an optimal control-based strategy to enhance the estimation of impulse-like disturbances in continuously monitored linear classical and quantum systems by exploiting non-equilibrium states. 
Using optimal estimation techniques for linear Gaussian systems to collect information from the temporal vicinity of the disturbance, we cast the minimization of disturbance estimation uncertainty as a nonlinear optimal control problem over time-dependent system parameters.
The resulting method dynamically shapes the estimation covariances through parametric modulation, maximizing information gain at a known impulse time. This differs fundamentally from conventional squeezing protocols using periodic modulation that effectively degrade inference of impulse-like disturbances.
Applied to nanomechanical resonators and levitated nanoparticles, optimal parametric driving reduces estimation variance by up to a factor of two relative to steady-state operation. 
%The framework provides a general approach to enhance precision in continuously monitored sensing platforms.%demonstrating that non-equilibrium preparation must be purpose-designed for jump detection rather than adapted from standard squeezing methods.

\end{abstract}


\end{frontmatter}
%===============================================================================

\section{Introduction}
Improving the sensitivity of measurement devices by exploiting non-equilibrium effects in their dynamics is an active research topic across various fields, such as nanoelectromechanical systems (NEMS), quantum optomechanics, trapped ions, and superconducting circuits (\cite{schmid2016fundamentals, kamba2025quantum, ge2019trapped, didier2014perfect}). Many of these systems are well described as stochastic linear dynamic systems (\cite{magrini_real-time_2021, jacobs2014quantum, rasmussen2021superconducting}) with dynamically adjustable parameters, e.g., by external modulation of  the system's resonance frequency or damping. 
%However, additional degrees of freedom often exist that can be utilized to manipulate their dynamical behavior, for instance through the external modulation of system parameters such as frequency or damping. 
It is well known that dynamic manipulation of these parameters can influence second-order statistics, which allows a significant enhancement of measurement sensitivity, e.g., using thermomechanic (\cite{rugar1991mechanical}) or quantum (\cite{cosco2021enhanced, wollman2015quantum}) squeezing. However, using such non-equilibrium effects requires to control the nonlinear behavior in these second-order statistics, which makes their analytical treatment challenging. %Modern tools from signal processing and control theory provide powerful methods to address these challenges effectively.

A particular class of sensing problems involves the detection and estimation of jump processes in continuous systems, in which the system undergoes an instantaneous change in magnitude at a discrete point in time due to an external disturbance (\cite{wang2023beating}). Examples include frequency jumps induced by nanoscale mass additions in NEMS resonators (\cite{ruz2020effect,sansa2020optomechanical}) or momentum kicks imparted to levitated nanoparticles through gas collisions (\cite{magrini_real-time_2021}, \cite{barker2024collision}). Recently, it has been shown that the disturbance causing such instantaneous changes can be effectively estimated from time traces in the vicinity of the disturbance event using Kalman–Bucy filtering and smoothing techniques, which are well suited for classical linear Gaussian dynamics (\cite{schmerling2025optimal}). However, it remains unclear how to (optimally) use non-equilibrium effects due to parametric driving and whether existing methods using squeezed states of motion are suitable for this specific scenario.

Hence, the present work extends methods introduced in (\cite{schmerling2025optimal}) to Gaussian open quantum systems while incorporating non-equilibrium effects to further enhance their sensitivity to impulse-like disturbances. To achieve this, we use the fact that for quantum Gaussian systems, their complete statistical behavior can be explicitly described by the Kalman–Bucy forward and backward filter equations. We then formulate the desired enhancement in the precision of a specific measurement quantity as a nonlinear optimal control problem, which can be efficiently solved using modern optimization techniques. Finally, we illustrate the proposed procedure to enhance the estimation of momentum kicks imparted on two state-of-the-art mechanical sensing devices: classical NEMS resonators and optically levitated nanoparticles in the quantum regime.

%\input{math_quantum_case}
%\section{Harmonic Oscillator}
%The Harmonic oscillator is a standard system in both classical and quantum mechanics because it is one of the few systems for which it is possible to derive analytic solutions to its equations of motion. Furthermore, its dynamics in the quantum case often resemble behavior which can be seen in the classical model as well. Therefore, it is a standard system to investigate the differences between classical and quantum systems.
\section{Open Quantum Systems} \label{chap:openquantharm}
%To derive the equations of motion of a quantum system, we have to set the stage by introducing the concept of conditional (stochastic) master equations, where we will focus on (\cite{jacobs2014quantum, milburn2011introduction}). 
The time evolution of a continuously monitored quantum system is given by the stochastic master equation (SME), which extends the Lindblad master equation for open quantum systems by incorporating continuous information gained through measurement and the corresponding measurement back-action into a conditional state described by the density matrix $\denmatr(t)$.
In the following we set $\hbar=1$ for simplicity. For a system with Hamiltonian operator $\hamil$ and a set of collapse operators $\measop_i$ corresponding to different interaction channels between the quantum system and its environment, the SME can be written as
\begin{align}
        \dif \denmatr(t) =&
            -\left(i\left[\hamil(t),\denmatr(t)\right] + \sum_{i=1}^{m} \sopdet[\measop_i(t)]\denmatr(t)\right) \dif t \nonumber\\
            &+  \sum_{i=1}^{m}\sqrt{\meffi_i(t)}\sopsto[\measop_i(t)]\denmatr(t)\dif W_i(t), \label{eqn:SME}
\end{align}
where the first term describes the deterministic evolution of the quantum state, and the second term accounts for the stochastic disturbance introduced by continuous measurements represented by the Wiener increments $\dif W_i$ with $\Exp{\dif W_i}=0$ and $(\dif W_i\dif W_j)=\delta_{ij}\dif t$. The superoperators acting on the density matrix $\sopdet[\measop]$ and $\sopsto[\measop]$ have the form 
\begin{equation}
    \begin{array}{cc}
         \sopdet[\measop]\denmatr&= \measop\denmatr\adj{\measop} - \frac{1}{2}(\adj{\measop}\measop\denmatr + \denmatr\adj{\measop}\measop)  \\
         \sopsto[\measop]\denmatr&=\measop\denmatr + \denmatr\adj{\measop} - \trace{\denmatr(\measop + \adj{\measop})} 
    \end{array}
\end{equation}
and account for the unobserved and observed interaction of each channel, see \cite{milburn2011introduction}.

The state $\denmatr(t)$ above is conditioned on the continuous (homodyne) measurement outcomes, i.e., 
\begin{equation} \label{eqn:smeInnovation}
    \dif Y_i(t) =  \sqrt{\meffi_i(t)}\trace{\denmatr(t)(\measop_i(t) + \adj{\measop}_i(t))}\dif t + \dif W_i(t),
\end{equation}
with the so-called measurement efficiencies $\meffi_i\in[0,1]$.
%The random disturbance processes $\dif W_i$ originating from continuously measuring the system, for example a homodyne measurement, are called innovations and can be decomposed according to
%\begin{equation} \label{eqn:smeInnovation}
%    \dif Y_i(t) =  \sqrt{\meffi_i(t)\mrate_i(t)}\trace{\denmatr(t)(\measop_i(t) + \adj{\measop}_i(t))}\dif t + \dif W_i(t)
%\end{equation}
%where $\dif Y_i$ is the differential increment of the measurement signal from the $i$-th measurement channel which hence acts back on the state. 
A key distinction between quantum and classical dynamics is that any extraction of information from a quantum system inherently disturbs that system, which imposes a fundamental limit on the achievable knowledge of its state. %This feature is captured by the parameters $\mrate_i$ and $\meffi_i$, which denote the measurement rate and efficiency, respectively. 
To describe this one can introduce the measurement rates $\mrate_i=||\measop||_o$ defined as some suitable operator norm $||.||_o$ of the collapse operators. These specify the speed at which information is extracted and, consequently, the strength with which the measurement disturbs the system. The efficiencies $\eta_i$ specify the fraction of extracted information that is accessible to the observer from each channel.

\section{Gaussian state Evolution}
Analytic solutions to stochastic master equations are generally intractable; however, one important class of exceptions is formed by systems with Hamiltonians at most quadratic of the form
\begin{equation} \label{eqn:quadHamil}
    \hamil(t) = \frac{1}{2}\trans{\quadop}\hamilMatr(t)\quadop + \trans{\quadop}\hamilVec(t)+\adj{\hamilVec}(t)\quadop,
\end{equation}
where the $2n$-dimensional vector $\quadop$ collects all canonical conjugate operator pairs of a system with $n$ bosonic modes, i.e., $\trans{\quadop} = (\trans{\hat{\mathbf{q}}}, \trans{\hat{\mathbf{p}}})$ satisfying the canonical commutation relation $[\quadop,\trans{\quadop}] = i\sympmatr$, where $\sympmatr$ denotes a symplectic matrix (\cite{ferraro2005gaussian}). Further, $\hamilVec(t)$ is a vector and $\hamilMatr(t)$ is a symmetric matrix of appropriate dimensions. 
%For open quantum system another requirement for the system to stay within the gaussian regime is that the 
Additionally, we assume that the collapse operators are linear operators of the form $\measop_i = \trans{\collapseVec}_i(t)\quadop$ where $\collapseVec_i(t)\in\Imaginaries^{2n}$. 
Such systems are populated by Gaussian states (see \cite{albarelli_pedagogical_2024}) which are fully described by a Gaussian phase space distribution (i.e., their Wigner function). Gaussian states can be fully characterized by their first and second statistical moments: the mean vector given by $\quadvec(t)=\trace{\quadop\denmatr(t)}$ and covariance matrix given by $\quadcov(t)=\trace{\frac{1}{2}\{(\quadop-\quadvec),\trans{(\quadop-\quadvec)}\}\denmatr(t)}$ with anticommutator $\{ \}$. 
%The mean of a canonical operator acting on the density matrix is given by $\quadvec(t)=\trace{\quadop\denmatr(t)}$, and its covariance matrix by $\quadcov(t)=\trace{\{(\quadop-\quadvec),\trans{(\quadop-\quadvec)}\}\denmatr(t)}$. 
The uncertainty principle is enforced through the Robertson–Schrödinger matrix inequality (\cite{simon1994quantum})
\begin{equation} \label{eqn:cannocial_com_rel}
    \quadcov + \frac{i}{2}\sympmatr \succeq 0.
\end{equation}

%As we assume that all states which populate our system are sufficiently described by their first and second quadrature moment we can simplify the general dynamics introduced in (\ref{eqn:SME}) to a set of linear stochastic and quadratic deterministic differential equations describing first and second order moments. 
We will replace $\state$ and $\cov$ with $\fwrdstate$ and $\fwrdcov$ to indicate that they are conditioned on past measurements and evolve forward in time. We furthermore omit all explicit time dependencies for readability. % it is emphasized that all variables and parameters might be time dependent.
Inserting the quadratic Hamiltonian (\ref{eqn:quadHamil}) and the linear collapse operators $\measop_i = \trans{\collapseVec}_i\quadop$ into the SME (\ref{eqn:SME}) and the measurement equation (\ref{eqn:smeInnovation}) yields the evolution of the mean
\begin{equation} \label{eqn:kalman_fwrd_mean}
    \dif\fwrdstate =  \dynA\fwrdstate\dif t + (\fwrdcov\trans{\dynC} - \trans{\dynN})\matrsqr{\dynEta}\dif \prcn,
\end{equation}
initial condition $\fwrdstate(0)=\fwrdstate_0$. Here $\dif \prcn$ is called the innovation processes that can be expressed using the differential measurements increment as $\dif\prcn = \dif\meas - \dynC\fwrdstate\dif t$. Analogously, the evolution of the covariance matrix obeys the well-known (deterministic) differential matrix Riccati equation 
\begin{equation} \label{eqn:fwrd_riccati}
    \ddt \fwrdcov = \dynA\fwrdcov + \fwrdcov\trans{\dynA} + \dynQ - (\fwrdcov\trans{\dynC} - \trans{\dynN})\dynEta\trans{(\fwrdcov\trans{\dynC} - \trans{\dynN})},
\end{equation}
which evolves forward in time with symmetric positive definite initial condition $\fwrdcov(0)=\fwrdcov_0 \succ \boldsymbol{0}$. The dynamics of the first and second moments above are related the SME description by the real matrices
\begin{equation} \label{eqn:dynMatrices}
    \begin{array}{cclccl}
         \dynA&=& \sympmatr(\hamilMatr + \imagPart{\collapseVec\adj{\collapseVec}}), & \dynC &=& 2\trans{\realPart{\collapseVec}}, \\
         \dynQ&=& \sympmatr \realPart{\collapseVec\adj{\collapseVec}}\trans{\sympmatr}, & \dynEta &=& \diag{\eta_1,\ldots,\eta_m}, \\
         \dynN &=& \sympmatr\imagPart{\collapseVec}, &&&
    \end{array}
\end{equation}
see also (\cite{zhang2017prediction}). 
%Here $\realPart{}$ and $\imagPart{}$ extract the real and imaginary parts of the corresponding expressions. 
These equations are of the same form as the classical Kalman filter, even though they were derived in a different setting (\cite{kalman_new_1961, edwards2005optimal}). Unlike in the classical case, the uncertainty principle in quantum systems due to \eqref{eqn:cannocial_com_rel} is enforced by the specific form of $\dynQ$, $\dynEta$ and $\dynA$ in \eqref{eqn:dynMatrices}.

One can also construct an operator $\hat{E}$ conditioned on measurement outcomes at future times called the effect operator, which can be interpreted as an effective  positive operator-valued measure (POVM) at a past time, see \cite{lammers2024quantum}.
The effect operator's evolution equations are given as the adjoint of the conditional state equation (\ref{eqn:SME})-(\ref{eqn:smeInnovation}), propagating backward in time with $\tau = T-t$ from some final time $T$. 
%The derivation can be done analogously to (\ref{eqn:SME})-(\ref{eqn:smeInnovation}) for the so called backwards effect operator as the adjoint of the conditional state by reversing time via $\tau = T - t$, starting from some final time $T$. In the physics literature this is referred to as state retrodiction (\cite{lammers2024quantum}). 
If we again make the assumptions (\ref{eqn:quadHamil})-(\ref{eqn:cannocial_com_rel}), the effect operator is also Gaussian and the evolution of its mean $\backstate(\tau)=\frac{\trace{\quadop \hat{E}(\tau)}}{\trace{\hat{E}(\tau)}}$ where $\hat{E}(\tau)$ is explicitly normalized as it does not necessarily preserve trace. Then its dynamics yields,
\begin{equation} \label{eqn:kalman_back_mean}
    \frac{\dif}{\dif \tau}\backstate =  -\dynA\backstate + (\backcov\trans{\dynC} + \trans{\dynN})\matrsqr{\dynEta}\dif \prcn,
\end{equation}
which is the analog to (\ref{eqn:kalman_fwrd_mean}) derived for reversed time with terminal condition $\backstate(0)=\backstate_0$ and
the backwards innovation process being $\dif \prcn = \dif\meas - \dynC\backstate \dif \tau$. Equivalently, the backwards evolution of the second-order moment becomes 
\begin{equation}\label{eqn:back_riccati}
    \frac{\dif}{\dif\tau} \backcov = -\dynA\backcov - \backcov\trans{\dynA} + \dynQ - (\backcov\trans{\dynC} + \trans{\dynN})\dynEta\trans{(\backcov\trans{\dynC} + \trans{\dynN})},
\end{equation}
with the covariance matrix of the backwards effect operator $\backcov(\tau) =\trace{\frac{1}{2}\{(\quadop-\backstate),\trans{(\quadop-\backstate)}\}\hat{E}(\tau)}$ with positive definite terminal condition $\backcov(T)=\backcov_T \succ \boldsymbol{0}$, cf.  \cite{zhang2017prediction}. This procedure is also referred to as state retrodiction or retrofiltering in the physics literature. 

By combining forward and backward filtering at some time instance $t_p$, one can arguably extend classical state smoothing into the quantum domain. However, this procedure naively yields state estimates that are not Hermitian in general and can violate the uncertainty principle, see \cite{guevara2015quantum, laverick2021quantum}. In our case, however, we aim to estimate an external Dirac-like disturbance, instantaneously changing the state at time $t_p$. We can thus use forward filtering and backward retrodiction techniques independently to estimate the states immediately before and after $t_p$.% as shown in %Fig.~\ref{fig:filter_retrodiction_illustration}.
%\begin{figure}[t]
%    \centering
%    \def\svgwidth{0.5\textwidth}
%    \input{figures/fwrd_back_filter_illustration.pdf_tex}
%    \vspace{-0.7cm}
%    \caption{Illustration of forward filtering and backward retrodiction for state estimation.}
%    \label{fig:filter_retrodiction_illustration}
%\end{figure}

\section{Optimal Impulse Estimation}
Suppose that a Dirac-like impulse $\alpha\boldsymbol{\delta}(t-t_p)$ with unknown magnitude $\alpha$ acts on the Gaussian quantum system at a known time $t_p$, as an external coupling, instantaneously changing its first moment $\quadvec$ by an offset $\Delta\quadvec$. We assume that $\Delta\quadvec$  maps linearly to the impulse magnitude with $\alpha\propto\trans{\impulseDirection}\Delta\quadvec$ whereby $\impulseDirection\in\Reals^{2n}$ with $||\impulseDirection||_2=1$ specifies the direction of the impulse along the canonical quadratures. For example, the magnitude of an external momentum kick is directly proportional to the momentum displacement of the observed system.

Building on the concepts of forward and backward filtering introduced above, we now address the optimal estimation of Dirac-like impulse disturbances from measurement data $\meastimetrace=\{\meas(t)|t\in[0,T]\}$ of a system described by (\ref{eqn:fwrd_riccati}) and (\ref{eqn:back_riccati}), where the dynamic model $\modeltimetrace = \{\dynA(\dynParams(t)),\dynC(\dynParams(t)),\dynQ(\dynParams(t)),\dynN(\dynParams(t)),\dynEta(\dynParams(t))|t\in[0,T]\}$ is known. Further, without loss of generality, we assume that all matrices can be expressed via the time-dependent vector of parameters $\dynParams(t)\in\paramsSet$. The set of admissible parameters $\paramsSet$ is given by the specifics of the system considered, e.g., constraints on experimentally variable parameters.

The discontinuity, breaks the temporal correlation of the system across $t_p$, which makes model (\ref{eqn:kalman_fwrd_mean})-(\ref{eqn:back_riccati}) invalid at that instant. Nevertheless, for the intervals $t\in[0,t_p)$ and $t=T-\tau\in[t_p,T]$, the model remains valid, allowing the change $\Delta\state$ to be estimated by separating the information before and after the impulse into two datasets:
\begin{equation*}
    \begin{array}{l}
         \Dataset_1=\{\meastimetrace,\modeltimetrace|t\in[0,t_p)\},  
         \\
         \Dataset_2=\{\meastimetrace,\modeltimetrace|t=T-\tau\in[t_p,T)\}. 
    \end{array}
\end{equation*}
The change in the first moment at $t_p$ is then given by
\begin{equation}
    \Delta\state(t_p) = \state(t_p)|\Dataset_1 - \state(t_p)|\Dataset_2,
\end{equation}
where $\state(t_p)|\Dataset_i$ denotes the first moment conditioned on dataset $\Dataset_i$. 

The conditioned first moment $\state(t_p)|\Dataset_1$ depends only on past data and is obtained from (\ref{eqn:kalman_fwrd_mean})–(\ref{eqn:fwrd_riccati}), which gives the forward mean $\fwrdstate_{\Dataset_1}(t)$ with covariance $\fwrdcov_{\Dataset_1}(t)$ for $t\in[0,t_p)$. Its limit as $t \to t_p$ provides the conditioned estimate, $\state(t_p)|\Dataset_1=\lim_{t\to t_p}\fwrdstate_{\Dataset_1}(t)$. The effect of the initial condition $\fwrdstate(0)$ and $\fwrdcov(0)$ decays exponentially and becomes negligible for sufficiently long data records.  

Conversely, $\state(t_p)|\Dataset_2$ depends only on future data and is obtained by solving the backward equations (\ref{eqn:kalman_back_mean}) and (\ref{eqn:back_riccati}) for $T-\tau\in[t_p,T]$, yielding the backward mean $\backstate_{\Dataset_2}(t)$ and covariance $\backcov_{\Dataset_2}(t)$, with $\state(t_p)|\Dataset_2=\backstate_{\Dataset_2}(t_p)$. Assuming no prior information about the impulse, the uncertainty in the estimate is fully determined by the forward and backward filters, and the error covariance of the impulse estimate is given by
\begin{equation} \label{eqn:impulseVariance}
    \cov_{\Delta\state}(t_p) = \fwrdcov_{\Dataset_1}(t_p) + \backcov_{\Dataset_2}(t_p),
\end{equation}
assuming that the forward and backward estimates are conditionally independent (\cite{schmerling2025optimal}).

\subsection{Covariance Minimization}
As previously noted, the system matrices in (\ref{eqn:fwrd_riccati}) and (\ref{eqn:back_riccati}) can be fully time dependent via some parameter vector $\dynParams(t)$. If these time-dependent parameters can be externally controlled, they provide additional degrees of freedom to enhance measurement sensitivity. For Gaussian states, maximizing the information about a quantity is equivalent to minimizing its variance. Therefore, to maximize the sensitivity with respect to the impulse's magnitude $\alpha$, we seek to minimize the combined forward and backward covariance at the impulse time $t_p$ in accordance with (\ref{eqn:impulseVariance}). This leads to the following optimization problem
\begin{equation} \label{eqn:continious_ocp}
    \begin{array}{ccl}
         \min\limits_{\dynParams(t)} && \trans{\impulseDirection}\cov_{\Delta\state}(t_p)\impulseDirection + \gamma_{\text{reg}}\int_{0}^{T}||\dynParams(t)||_2\dif t\\
         \text{s.t.} & \ddt \fwrdcov &= \mathbf{f}(\fwrdcov,\dynParams(t)) \quad \text{from } \eqref{eqn:fwrd_riccati}, \\
                     & \ddtau \backcov &= \mathbf{g}(\backcov,\dynParams(\tau)) \quad \text{from } \eqref{eqn:back_riccati},\\
                     && \dynParams \in \paramsSet,
    \end{array}
\end{equation}
which minimizes the impulse estimation uncertainty along the direction $\impulseDirection$ of the canonical quadrature variables at time $t_p$. The covariance matrices are constrained by the Riccati dynamics of their respective filters, while the model parameters $\dynParams$ are restricted to the set of admissible parameters $\paramsSet$, and the cost function is extended with a regularization term with the norm $||.||_2$ in the parameters and $\gamma_{\text{reg}}\in\Reals^{+}$. Conceptually, this enhances sensitivity by shaping the covariance ellipse of the quantum state along the direction of the measured quadrature. 

\begin{figure}[t]
    \centering
    \def\svgwidth{0.5\textwidth}
    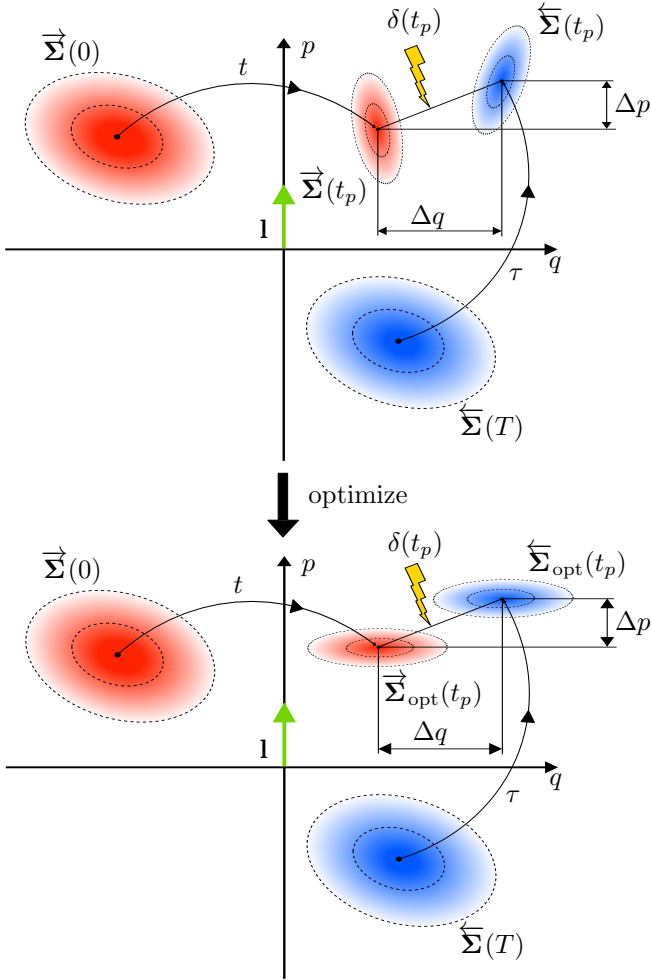
    \vspace{-0.8cm}
    \caption{Illustration of the impulse estimation improvement by optimizing the covariance dynamics. Top: unoptimized case. Bottom: optimized case with minimized variance along the $p$ direction, e.g., for the case of a momentum kick.}
    \label{fig:pahse_space_opt}
\end{figure}

This concept is illustrated in Fig.~\ref{fig:pahse_space_opt}. Since the Riccati equations (\ref{eqn:fwrd_riccati}) and (\ref{eqn:back_riccati}) are nonlinear in the covariances, the optimization problem is generally nonlinear and may be nonconvex, necessitating the use of numerical tools for non-convex optimization.

\section{Numerical Examples}
In the following section, we investigate the estimation of momentum kicks on two representative systems commonly encountered in ultra-sensitive mechanical experiments: nanomechanical resonators (\cite{rugar1991mechanical}), limited by thermomechanical noise and optically levitated nanoparticles (\cite{magrini_real-time_2021}) in the quantum regime. For these two simple systems it is sufficient to consider a single mode with single canonical quadrature pair $\quadop = \trans{[\hat{q},\hat{p}]}$. Further, we consider momentum kicks as impulse-like disturbances, i.e., $\impulseDirection=\trans{[0 \quad 1]}$.
Since numerical and experimental ensemble statistics of $\Delta\state$ aligned exceptionally well with their theoretical covariances $\cov_{\Delta\quadvec}(t_p)$ from \eqref{eqn:impulseVariance} (see \cite{schmerling2025optimal}), we omit individual stochastic trajectories due to space limitations.

\subsubsection{Nanomechanical Resonators}
Thermomechanical noise squeezing in NEMS resonators is a well-understood phenomenon that has been exploited to enhance the sensitivity of force measurements by dynamically shaping the noise statistics (\cite{rugar1991mechanical}). A simple method to manipulate the second statistical moments of a mechanical oscillator is to vary its harmonic frequency $\Omega(t)$ by changing its spring stiffness. Defining a time-varying frequency with a constant base frequency $\Omega_0$ and a dimensionless modulation parameter $p(t)$ as $\Omega(t) = \Omega_0(1 + p(t))$. Furthermore, viscous damping acts on the system via the damping rate $\Gamma$. We can rewrite the time-varying model of a classical NEMS resonator in the quantum Gaussian framework (\ref{eqn:kalman_fwrd_mean})-(\ref{eqn:dynMatrices}) using
\begin{equation}
\begin{array}{clcl}
     \dynA=&  \begin{bmatrix}
        0 & \Omega_0 \\
        -\Omega_0(1+p(t)) & -\Gamma
    \end{bmatrix}, & \dynC=& \begin{bmatrix}
        \frac{q_{\text{zpf}}}{\sqrt{\mnoise}} & 0 
    \end{bmatrix},
    \\
    \dynQ=& 
    \begin{bmatrix}
        0 & 0 \\
        0 & \frac{\fnoise}{p_{\text{zpf}}^2}
    \end{bmatrix},
    &
    \dynEta =& 1, \quad \dynN = 0,
\end{array}
\end{equation}
where the quadratures are scaled such that both are unit-less. We further assume that the NEMS resonator resides in a high-temperature thermal equilibrium with the surrounding bath as the dominant noise source, which acts on the resonator by a random force. Hence, the force noise and zero-point motions are given by
\begin{equation}
\begin{array}{clclcl}
     \fnoise&=\frac{4k_{\text{B}}\Theta\Gamma}{m},& q_{\text{zpf}} &= \sqrt{\frac{\hbar}{m\Omega_0}},& p_{\text{zpf}} &= \sqrt{\hbar m\Omega_0}
\end{array}
\end{equation}
as defined by the fluctuation-dissipation theorem (\cite{callen1951irreversibility}) for the high temperature regime ($k_{\text{B}} \Theta\gg \hbar \Omega$), where $\Theta$ denotes the bath temperature and $k_{\text{B}}$ is the Boltzmann constant and $\hbar$ is the reduced Planck's constant. The measurement noise PSD $\mnoise$ is given by the photon shot noise of an interferometric measurement (\cite{rugar1991mechanical}).

\subsubsection{Levitated Nanoparticle}
The next example concerns a nanosphere confined in a harmonic potential formed by an optical tweezer, i.e., a focused laser beam. 
%In this setup, a silica sphere with a diameter of several hundred nanometers is trapped by the radiation pressure of a focused laser beam. 
The interaction between the particle and the light field encodes information about the particle’s position as phase modulation of the scattered light, which can be extracted via homodyne detection, while also introducing back-action noise to the particle. The system is well described as an open quantum harmonic oscillator under linear Gaussian measurement, and can therefore be analyzed using our framework. The trapping frequency is proportional to the square root of the laser power, $\Omega(t)\propto \sqrt{P(t)}$. Furthermore, the PSD of the back-action noise resulting from photon recoil is proportional to the laser power via the measurement rate constant $\kappa\propto P(t)$. Again, introducing the modulation as $P(t)=P_0(1 + p(t))$, the quantum-optical system takes the form (\cite{magrini_real-time_2021, jacobs2014quantum}) 
\begin{equation} 
\begin{array}{clcl}
     \dynA=&  \begin{bmatrix}
        0 & \Omega_0 \\
        -\Omega_0\sqrt{(1+p(t))} & -\Gamma
    \end{bmatrix}, & \dynC=& \begin{bmatrix}
        \sqrt{8\kappa_0(1+p(t))} & 0 
    \end{bmatrix},
    \\
    \dynQ=& 
    \begin{bmatrix}
        0 & 0 \\
        0 & 2\kappa_0(1+p(t))
    \end{bmatrix},
    &
    \dynEta =& \eta_{\text{hom}}, \quad \dynN = 0,\label{eqn:statespaceparticle}
\end{array}
\end{equation}
where the parameter $\kappa_0$ is the steady state measurement rate for $p=0$. As the trapping laser simultaneously provides the confining potential and enables position readout with measurement efficiency $\eta_{\text{hom}}$, its power determines both the measurement strength and the associated back-action noise. The measurement noise is given by photon shot noise, which itself scales with laser power, see e.g., \cite{magrini_real-time_2021}. %Thus, decreasing the measurement noise. 
%However increasing the power also leads to more force back-action which increases the force noise.

\subsubsection{Solution of the Optimal Control Problem}
Based on (\ref{eqn:continious_ocp}), we approximate the continuous-time optimal control problem (OCP) by a discrete-time formulation that can be solved using standard non-convex optimization solvers such as IPOPT (\cite{wachter2006implementation}), implemented in the optimal control framework CASADI (\cite{Andersson2019}). We use multiple shooting with piecewise-constant control inputs $p_k$ and explicit fourth  order Runge-Kutta schemes for forward and backward numerical integration of the differential Riccati equations (\ref{eqn:fwrd_riccati}) and (\ref{eqn:back_riccati}).
\subsubsection{Results}
\begin{figure}[ht!]
    \centering
    \includegraphics[width=0.475\textwidth]{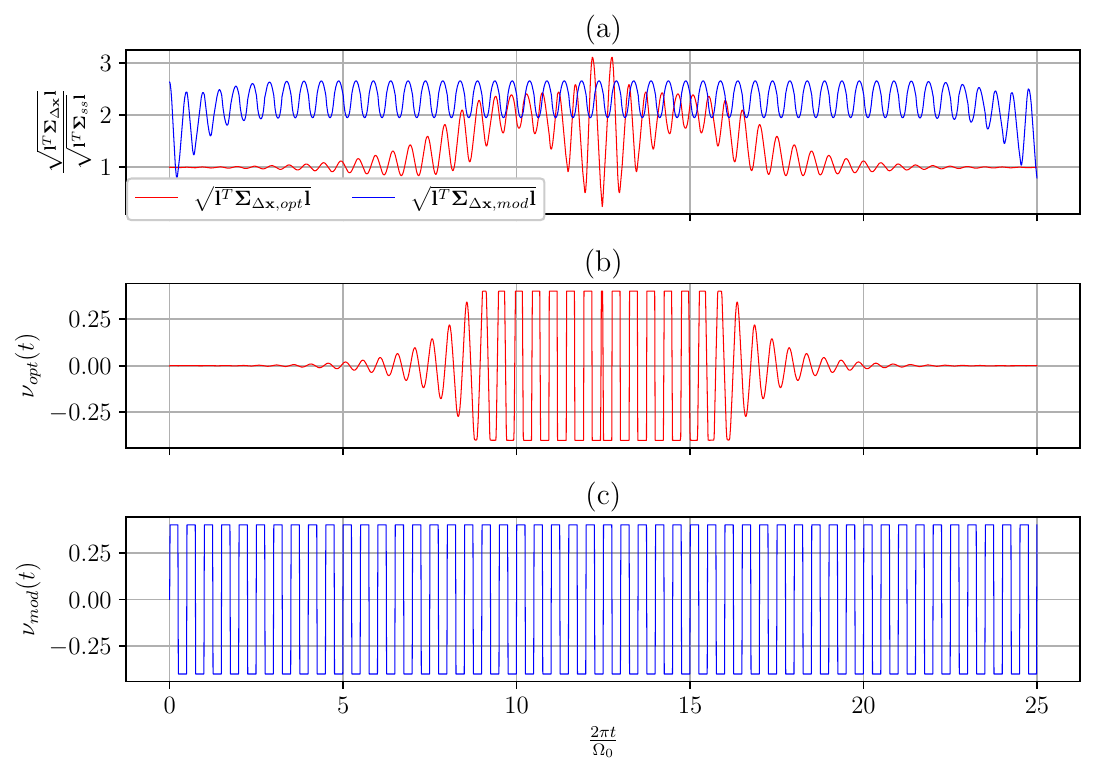}
    \vspace{-0.5cm}
    \caption{(a) Momentum uncertainty of the NEMS resonator. Red line: square root of the sum of forward and backward covariances with optimized modulation signal. Blue line: square root of the sum of forward and backward covariances for a rectangular modulation signal. (b) Optimized modulation signal. (c) Rectangular modulation with $2\Omega_0$ frequency.}
    \label{fig:nems_opt_squeez}
\end{figure}
Figure \ref{fig:nems_opt_squeez} compares the relative square root of the sum of forward and backward variances in the NEMS system for the optimized modulation parameter against a simple rectangular modulation at twice the mechanical resonance frequency known to create maximum squeezing (\cite{asjad2014robust}). Initial and terminal values of the forward and backward Riccati equations were set to their steady-state solution, i.e., $\fwrdcov(0) = \fwrdcov_{\text{ss}}$ and $\backcov(0) = \backcov_{\text{ss}}$. The physical parameters of the simulated system  are taken from \cite{rugar1991mechanical}: angular frequency $\frac{\Omega_0}{2\pi} = \SI{33.7}{kHz}$, damping rate $\Gamma = \SI{2.07}{Hz}$, and mass $m = \SI{2.8E-12}{kg}$ at a temperature of $\Theta=\SI{295}{K}$, resulting in a force noise PSD $\fnoise = \SI{5.3E-31}{N^2/Hz}$. The measurement noise floor is assumed to have a PSD $\mnoise = \SI{4E-28}{m^2/Hz}$. The modulation depth is constrained by $-0.4\leq p\leq0.4$. The optimized modulation signal is shown in Fig.~\ref{fig:nems_opt_squeez}b, and a rectangular reference modulation in Fig.~\ref{fig:nems_opt_squeez}c. Due to the regularization term in the optimization problem \eqref{eqn:continious_ocp}, the optimized protocol only modifies $p(t)$ in the vicinity of $t_p$ relevant to extract information. From Fig.~\ref{fig:nems_opt_squeez}a, it is evident that both modulation strategies affect the impulse estimation uncertainty $\trans{\impulseDirection}\cov_{\Delta\quadvec}\impulseDirection$. For the optimized modulation, the uncertainty initially increases from the steady-state value, but then decreases and reaches a minimum of approximately $0.49$ times the steady-state uncertainty at the center of the modulation interval. In contrast, the simple rectangular modulation results in a transient increase, followed by a limit-cycle oscillation around $2$ to $2.5$ times the steady-state uncertainty. The generation of mechanically squeezed states of motion via periodic modulation does \emph{not} enhance sensitivity for \emph{impulse-like disturbances} but even deteriorates performance.
\begin{figure}[t!]
    \centering
    \includegraphics[width=0.475\textwidth]{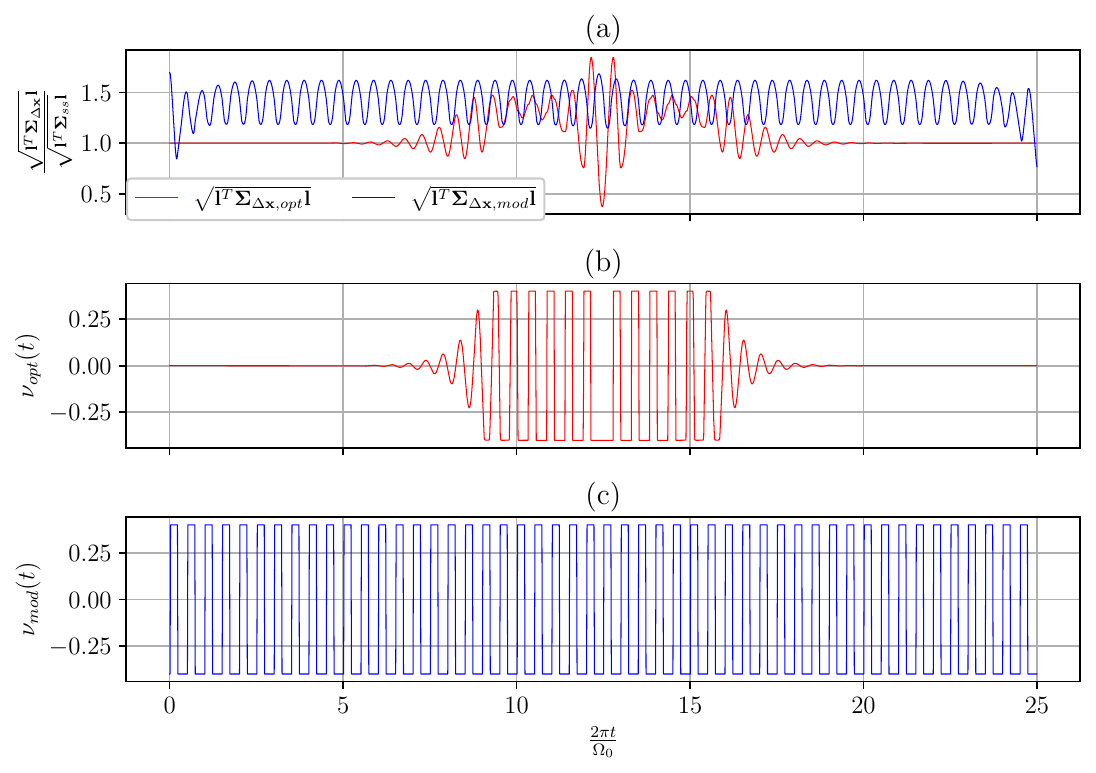}
    \vspace{-0.5cm}
    \caption{(a) Momentum uncertainty of a levitated nanoparticle. Red line: square root of the sum of forward and backward covariances with optimized modulation signal. Blue line: square root of the sum of forward and backward covariances for a rectangular modulation signal. (b) Optimized modulation signal. (c) Rectangular modulation with $2\Omega_0$ frequency and adjusted switching times.}
    \label{fig:lev_par_opt_squuez}
\end{figure}
The results for the levitated particle are shown in Fig.~\ref{fig:lev_par_opt_squuez}. The evolution of $\trans{\impulseDirection}\cov_{\Delta\quadvec}\impulseDirection$ for the optimized and the simple rectangular modulation protocols is shown in Fig.~\ref{fig:lev_par_opt_squuez}a, with the optimized modulation $p(t)$ in Fig.~\ref{fig:lev_par_opt_squuez}b. Figure~\ref{fig:lev_par_opt_squuez}c shows a rectangular reference modulation signal with adjusted switching times to account for the square root scaling from power to angular frequency. The parameters of the simulated system are taken from (\cite{magrini_real-time_2021}): $\frac{\Omega_0}{2\pi} = \SI{104}{kHz}$, damping rate $\Gamma = \SI{0.64}{Hz}$, and mass $m = \SI{4.5E-18}{kg}$. The measurement rate is $\kappa_0=\SI{41}{kHz}$ with a measurement efficiency $\eta_{\text{hom}} = 0.4$. The modulated uncertainty in Fig.~\ref{fig:lev_par_opt_squuez} shows qualitatively similar behavior to Fig.~\ref{fig:nems_opt_squeez}, with two notable differences: The modulation exhibits slightly shifted switching times and the modulation pattern at the center of the interval is different.

\begin{figure}[t!]
    \centering
    \includegraphics[width=0.475\textwidth]{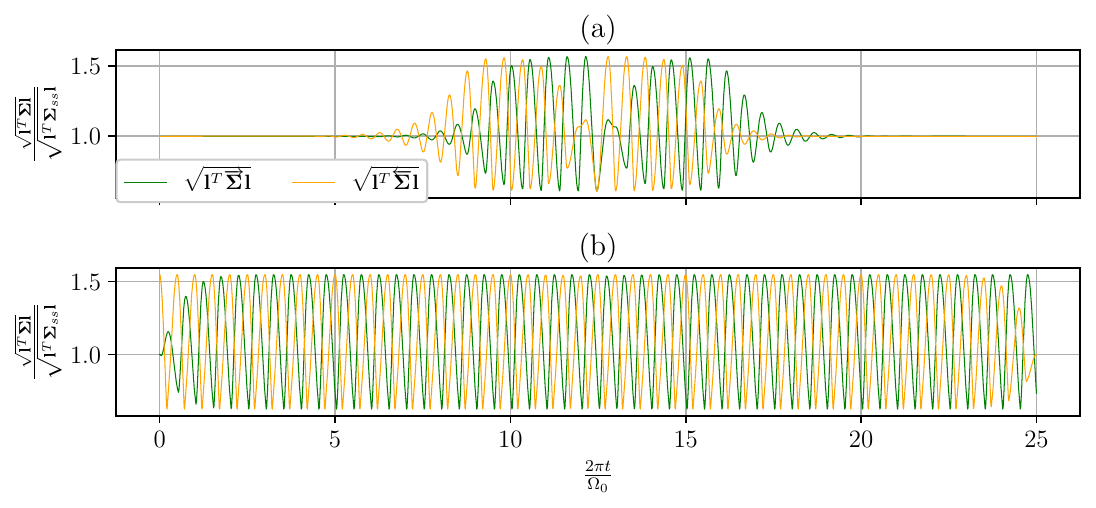}
    \vspace{-0.5cm}
    \caption{(a) Forward and backward momentum impulse estimation uncertainty  for optimized protocol. (b) Momentum impulse estimation uncertainty with the simple modulation protocol.}
    \label{fig:fwrd_back_cov_compare}
\end{figure}

These results show that optimal filtering approaches combined with modulation based on optimal control can be used to design protocols that enhance the sensitivity of continuously observed classical and quantum systems for detecting impulse-like disturbances. The resulting non-equilibrium state preparation strategies indeed achieve an estimation uncertainty significantly below their equilibrium value. It is evident that the simple modulation scheme showed in Fig.~\ref{fig:nems_opt_squeez}c and Fig.~\ref{fig:lev_par_opt_squuez}c, typically used in the literature, \emph{increase} rather than decrease the estimation uncertainty \emph{for impulse-like disturbances} in both scenarios. The reason for this can be found in Fig.~\ref{fig:fwrd_back_cov_compare}, which shows the time evolution of the normalized square root of forward and backward variance for the levitated nanoparticle using the optimized modulation protocol (a) and the simple rectangular modulation (b), respectively. Using simple modulation schemes results in a phase shift between $\trans{\impulseDirection}\fwrdcov\impulseDirection$ and $\trans{\impulseDirection}\backcov\impulseDirection$, leading to an overall increase in $\trans{\impulseDirection}\cov_{\Delta\quadvec}\impulseDirection$. In other words, whenever one gains additional information from the forward filter, one looses even more information in retrodiction and vice versa. The optimization-based protocol, on the other hand, accounts for this trade-off and still provides a significant reduction in uncertainty.

\section{Conclusion}

This paper developed a method to find non-equilibrium protocols for open Gaussian quantum systems that maximize the information about impulse-like disturbances accessible via continuous measurements through tailored parametric modulation. 
By coordinating forward filtering and backward retrodiction, the proposed control strategy minimizes the combined estimation variance at the impulse time and outperforms traditional squeezing schemes. In contrast to these approaches, the optimal non-equilibrium solution is explicitly tailored to a single inference time as required for impulse-like disturbances. 
% Missing: application example of recoil-based mass spectrometry

\subsubsection*{Declaration of generative AI and AI-assisted technology in the writing process:}
During the preparation of this work, the authors used ChatGPT in order to improve readability and
language. After using this tool/service, the author(s) reviewed and edited the content as needed and take full responsibility for the content of the publication.

%\begin{ack}
%Place acknowledgments here.
%\end{ack}

%\section*{DECLARATION OF GENERATIVE AI AND AI-ASSISTED TECHNOLOGIES IN THE WRITING PROCESS}
%During the preparation of this work the author(s) used [NAME TOOL / SERVICE] in order to [REASON]. After using this tool/service, the author(s) reviewed and edited the content as needed and take(s) full responsibility for the content of the publication.

\bibliography{literature}

@article{magrini_real-time_2021,
	title={Real-time optimal quantum control of mechanical motion at room temperature},
    author={Magrini, Lorenzo and Rosenzweig, Philipp and Bach, Constanze and Deutschmann-Olek, Andreas and Hofer, Sebastian G and Hong, Sungkun and Kiesel, Nikolai and Kugi, Andreas and Aspelmeyer, Markus},
    journal={Nature},
    volume={595},
    number={7867},
    pages={373--377},
    year={2021},
    publisher={Nature Publishing Group UK London}
}

@article{kalman_new_1961,
	title = {New Results in Linear Filtering and Prediction Theory},
	volume = {83},
	issn = {0021-9223},
	pages = {95--108},
	number = {1},
	journal = {Journal of Basic Engineering},
	author = {Kalman, R. E. and Bucy, R. S.},
	urldate = {2024-05-13},
	date = {1961-03-01},
	langid = {english},
    year = {1961},
}

@article{albarelli_pedagogical_2024,
	title = {A pedagogical introduction to continuously monitored quantum systems and measurement-based feedback},
	volume = {494},
	pages = {129260},
	journal = {Physics Letters A},
	author = {Albarelli, Francesco and Genoni, Marco G.},
	urldate = {2025-09-12},
	date = {2024-01-15},
    year = {2024},
}

@article{schmerling2025optimal,
  title={Optimal sensing of momentum kicks with a feedback-controlled nanomechanical resonator},
  author={Schmerling, Kaspar and Be{\v{s}}i{\'c}, Hajrudin and Kugi, Andreas and Schmid, Silvan and Deutschmann-Olek, Andreas},
  journal={Physical Review Applied},
  volume={23},
  number={5},
  pages={054016},
  year={2025},
  publisher={APS}
}

@article{ferraro2005gaussian,
  title={Gaussian states in continuous variable quantum information},
  author={Ferraro, Alessandro and Olivares, Stefano and Paris, Matteo GA},
  journal={arXiv preprint quant-ph/0503237},
  year={2005}
}

@article{simon1994quantum,
  title={Quantum-noise matrix for multimode systems: U (n) invariance, squeezing, and normal forms},
  author={Simon, Rajiah and Mukunda, Narasimhaiengar and Dutta, Biswadeb},
  journal={Physical Review A},
  volume={49},
  number={3},
  pages={1567},
  year={1994},
  publisher={APS}
}

@article{guevara2015quantum,
  title={Quantum state smoothing},
  author={Guevara, Ivonne and Wiseman, Howard},
  journal={Physical review letters},
  volume={115},
  number={18},
  pages={180407},
  year={2015},
  publisher={APS}
}

@article{rugar1991mechanical,
  title={Mechanical parametric amplification and thermomechanical noise squeezing},
  author={Rugar, D and Gr{\"u}tter, P},
  journal={Physical Review Letters},
  volume={67},
  number={6},
  pages={699},
  year={1991},
  publisher={APS}
}

@article{milburn2011introduction,
  title={An introduction to quantum optomechanics},
  author={Milburn, GJ and Woolley, MJ},
  journal={acta physica slovaca},
  volume={61},
  number={5},
  pages={483--601},
  year={2011},
  publisher={Slovenska Akademia Vied, Fyzikalny Ustav, Versita Sp. z o. o., Solipska 14 A~…}
}

@article{kamba2025quantum,
  title={Quantum squeezing of a levitated nanomechanical oscillator},
  author={Kamba, Mitsuyoshi and Hara, Naoki and Aikawa, Kiyotaka},
  journal={Science},
  volume={389},
  number={6766},
  pages={1225--1228},
  year={2025},
  publisher={American Association for the Advancement of Science}
}

@article{edwards2005optimal,
  title={Optimal quantum filtering and quantum feedback control},
  author={Edwards, Simon C and Belavkin, Viacheslav P},
  journal={arXiv preprint quant-ph/0506018},
  year={2005}
}

@article{cosco2021enhanced,
  title={Enhanced force sensitivity and entanglement in periodically driven optomechanics},
  author={Cosco, F and Pedernales, JS and Plenio, Martin B},
  journal={Physical Review A},
  volume={103},
  number={6},
  pages={L061501},
  year={2021},
  publisher={APS}
}

@book{jacobs2014quantum,
  title={Quantum measurement theory and its applications},
  author={Jacobs, Kurt},
  year={2014},
  publisher={Cambridge University Press}
}

@article{wachter2006implementation,
  title={On the implementation of an interior-point filter line-search algorithm for large-scale nonlinear programming},
  author={W{\"a}chter, Andreas and Biegler, Lorenz T},
  journal={Mathematical programming},
  volume={106},
  number={1},
  pages={25--57},
  year={2006},
  publisher={Springer}
}

@article{Andersson2019,
  author = {Joel A E Andersson and Joris Gillis and Greg Horn
            and James B Rawlings and Moritz Diehl},
  title = {{CasADi} -- {A} software framework for nonlinear optimization
           and optimal control},
  journal = {Mathematical Programming Computation},
  volume = {11},
  number = {1},
  pages = {1--36},
  year = {2019},
  publisher = {Springer},
}

@article{callen1951irreversibility,
  title={Irreversibility and generalized noise},
  author={Callen, Herbert B and Welton, Theodore A},
  journal={Physical Review},
  volume={83},
  number={1},
  pages={34},
  year={1951},
  publisher={APS}
}

@article{wollman2015quantum,
  title={Quantum squeezing of motion in a mechanical resonator},
  author={Wollman, Emma Edwina and Lei, CU and Weinstein, AJ and Suh, J and Kronwald, A and Marquardt, F and Clerk, Aashish A and Schwab, KC},
  journal={Science},
  volume={349},
  number={6251},
  pages={952--955},
  year={2015},
  publisher={American Association for the Advancement of Science}
}

@article{wang2023beating,
  title={Beating thermal noise in a dynamic signal measurement by a nanofabricated cavity optomechanical sensor},
  author={Wang, Mingkang and Perez-Morelo, Diego J and Ramer, Georg and Pavlidis, Georges and Schwartz, Jeffrey J and Yu, Liya and Ilic, Robert and Centrone, Andrea and Aksyuk, Vladimir A},
  journal={Science Advances},
  volume={9},
  number={11},
  pages={eadf7595},
  year={2023},
  publisher={American Association for the Advancement of Science}
}

@article{sansa2020optomechanical,
  title={Optomechanical mass spectrometry},
  author={Sansa, Marc and Defoort, Martial and Brenac, Ariel and Hermouet, Maxime and Banniard, Louise and Fafin, Alexandre and Gely, Marc and Masselon, Christophe and Favero, Ivan and Jourdan, Guillaume and others},
  journal={Nature Communications},
  volume={11},
  number={1},
  pages={3781},
  year={2020},
  publisher={Nature Publishing Group UK London}
}

@article{ruz2020effect,
  title={Effect of particle adsorption on the eigenfrequencies of nano-mechanical resonators},
  author={Ruz, JJ and Malvar, O and Gil-Santos, E and Calleja, M and Tamayo, J},
  journal={Journal of Applied Physics},
  volume={128},
  number={10},
  year={2020},
  publisher={AIP Publishing}
}

@book{schmid2016fundamentals,
  title={Fundamentals of nanomechanical resonators},
  author={Schmid, Silvan and Villanueva, Luis Guillermo and Roukes, Michael Lee},
  volume={49},
  year={2016},
  publisher={Springer}
}

@article{rasmussen2021superconducting,
  title={Superconducting circuit companion—an introduction with worked examples},
  author={Rasmussen, Stig Elkj{\ae}r and Christensen, Kasper Sangild and Pedersen, Simon Panyella and Kristensen, Lasse Bj{\o}rn and B{\ae}kkegaard, Thomas and Loft, Niels Jakob S{\o}e and Zinner, Nikolaj Thomas},
  journal={PRX Quantum},
  volume={2},
  number={4},
  pages={040204},
  year={2021},
  publisher={APS}
}

@article{ge2019trapped,
  title={Trapped ion quantum information processing with squeezed phonons},
  author={Ge, Wenchao and Sawyer, Brian C and Britton, Joseph W and Jacobs, Kurt and Bollinger, John J and Foss-Feig, Michael},
  journal={Physical review letters},
  volume={122},
  number={3},
  pages={030501},
  year={2019},
  publisher={APS}
}

@article{didier2014perfect,
  title={Perfect squeezing by damping modulation in circuit quantum electrodynamics},
  author={Didier, Nicolas and Qassemi, Farzad and Blais, Alexandre},
  journal={Physical Review A},
  volume={89},
  number={1},
  pages={013820},
  year={2014},
  publisher={APS}
}

@article{laverick2021quantum,
  title={Quantum Rauch-Tung-Striebel smoothed state},
  author={Laverick, Kiarn T},
  journal={Physical Review Research},
  volume={3},
  number={3},
  pages={033196},
  year={2021},
  publisher={APS}
}

@article{lammers2024quantum,
  title={Quantum retrodiction in Gaussian systems and applications in optomechanics},
  author={Lammers, Jonas and Hammerer, Klemens},
  journal={Frontiers in Quantum Science and Technology},
  volume={2},
  pages={1294905},
  year={2024},
  publisher={Frontiers Media SA}
}

@article{zhang2017prediction,
  title={Prediction and retrodiction with continuously monitored Gaussian states},
  author={Zhang, Jinglei and M{\o}lmer, Klaus},
  journal={Physical Review A},
  volume={96},
  number={6},
  pages={062131},
  year={2017},
  publisher={APS}
}

@article{barker2024collision,
  title={Collision-resolved pressure sensing},
  author={Barker, Daniel S and Carney, Daniel and LeBrun, Thomas W and Moore, David C and Taylor, Jacob M},
  journal={Physical Review A},
  volume={109},
  number={4},
  pages={042616},
  year={2024},
  publisher={APS}
}

@article{asjad2014robust,
  title={Robust stationary mechanical squeezing in a kicked quadratic optomechanical system},
  author={Asjad, Muhammad and Agarwal, GS and Kim, MS and Tombesi, Paolo and Giuseppe, G Di and Vitali, David},
  journal={Physical Review A},
  volume={89},
  number={2},
  pages={023849},
  year={2014},
  publisher={APS}
}
%\printbibliography
                                                   
%\appendix
%\section{A summary of Latin grammar}    % Each appendix must have a short title.
%\section{Some Latin vocabulary}              % Sections and subsections are supported  
                                                                         % in the appendices.
\end{document}